\documentclass{PoS}
\usepackage{graphicx}

\title{What are the degrees of freedom in the partonic fluid at RHIC ?}

\ShortTitle{Degrees of freedom in partonic fluid}

\author{Rene Bellwied\\
        Physics, Department, Wayne State University,\\
        Detroit, MI 48202, U.S.A.\\
        E-mail: \email{bellwied@physics.wayne.edu}}

\abstract{Recent RHIC data show evidence of multiple hadron
production mechanisms in heavy ion collisions compared to simple
fragmentation in vacuum. I will review the measurements of
collective flow, high momentum quenching, and two particle angular
correlations to show that neither thermal production nor string
fragmentation can describe the abundances, the angular distributions
or the kinematic properties of all hadrons produced at RHIC. The
proposed new hadronization mechanisms not only serve as evidence for
a deconfined partonic phase of matter, but also for strong coupling
of the degrees of freedom in the deconfined phase. I will point out
a surprising lack of flavor dependence in these properties at RHIC,
though, which might have to lead to further revisions of our
understanding of the relevant degrees of freedom in the partonic
phase and during the hadronization process.}

\FullConference{Critical Point and Onset of Deconfinement
          4th International Workshop\\
         July 9-13 2007\\
         GSI Darmstadt,Germany}

\begin{document}

\section{Introduction}\label{intro}

The measurements of anisotropic particle flow and jet quenching at
RHIC have revealed a deconfined state of matter at high temperature
and partonic density, which is characterized best as a near perfect
fluid, i.e. a collective state with an extremely low ratio of shear
viscosity to entropy. Indeed over the past year the experiments at
RHIC were able to experimentally verify the original conjecture of a
state near the quantum limit through several independent
measurements of the $\eta$/s ratio. Fig.1 shows a summary of
calculations based on $<$p$_{T}$$>$-fluctuation, light and heavy
quark elliptic flow, and quenching measurements \cite{gavin,
romatschke, lacey, phenix-vis}. These calculations are still model
dependent, but it is intriguing to recognize that, if the initial
conditions assumed in the models are correct, the new state of
matter is not well described by either perturbative QCD or a purely
hadronic model.

\begin{figure} [hbtl]
\hspace{3.cm}
\includegraphics[width=3.in,bb= 0 0 400 400]{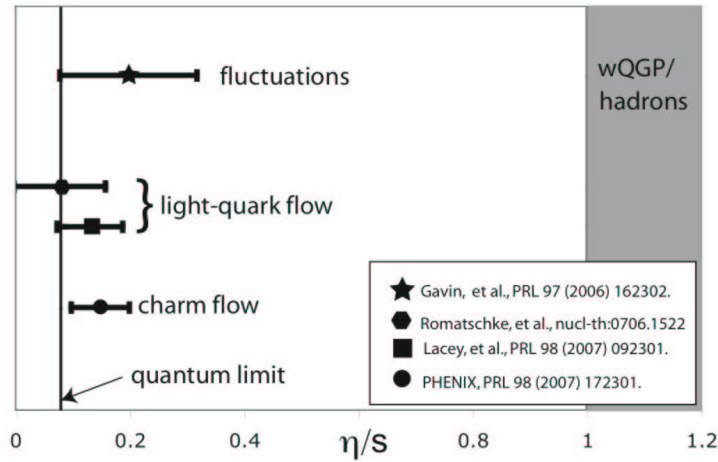} \label{fig:1}
\caption{Summary of recent model dependent determinations of the
shear viscosity over entropy ratio ($\eta$/s) based on measurements
from STAR and PHENIX.}
\end{figure}

This led to the definition of the sQGP, a strongly coupled Quark
Gluon Plasma, rather than the weak coupling phase we expected from
lattice QCD at sufficiently high initial temperature. Since then
recent lattice calculations and their comparison to hard thermal
loop calculations have revealed that the conditions at RHIC are not
sufficient to reach the weakly coupled limit, but that the LHC
energies might be sufficient to generate perturbative QCD like
conditions \cite{lattice}. This unexpected behavior of the
deconfined matter at RHIC poses the question of the nature of the
degrees of freedom in the partonic liquid and the details of the QCD
phase transition from a strongly coupled partonic system to the
hadronic gas. Clearly this mechanism might be different to the
fragmentation picture in vacuum which has been verified through
numerous measurements in elementary particle collisions over the
past three decades. In this paper I will try to show that the
particle emission spectra, the particle correlations as well as the
anisotropic particle flow reveals details of competing hadronization
mechanisms in heavy ion collisions which are either non-existent or
much less prominent in elementary collisions. These novel mechanisms
might have a profound impact on our understanding of the formation
of baryonic matter in the universe.

\section{The difference between hadronization in vacuum and in
medium}

In 1977 two competing papers appeared that tried to model the
hadronization process in high energy elementary collisions. The
initial Feynman/Fields paper describes the hadronization process in
vacuum through jet fragmentation \cite{feynman}. This approach was
later on extended to string fragmentation. Although no explicit
hadronization mechanism is given in this picture the hadronic
particle distribution can be parametrized through the fragmentation
function D$_{q}^{h}$, which yields the probability that a certain
parton 'q' fragments into a certain hadron 'h'. Baryon formation in
such a model generally requires the formation of a di-quark
structure, as a remnant of the initial hard scattering in a
proton-proton collision.

The other paper was by Das and Hwa \cite{hwa} and it tried to
describe hadronization through recombination or coalescence of
independent free quarks. The clustering of quarks is modeled through
a momentum overlap probability function. Again, there is no explicit
hadronization mechanism, but the particle emission spectra are
described well by this approach. This model has been the basis for
many recent recombination models used in heavy ion collisions (e.g.
\cite{bass,greco}).

Over the years the fragmentation approach has been widely accepted
as the main hadronization mechanism in vacuum, but recent evidence
in particle spectra measured in heavy ion collisions at RHIC has
re-ignited the interest in the recombination approach.

This is mainly due to two key heavy ion results, the particle
identified elliptic flow and quenching measurements. In both cases
the intermediate transverse momentum region of the measured spectra
has revealed a scaling, the so-called constituent quark scaling,
which can be interpreted as evidence for not only deconfinement but
also quark recombination. Figs. 2 and 3 show a summary of the
results for R$_{CP}$, the nuclear suppression factor defined as the
ratio of the transverse momentum spectra measured in different
centrality bins and scaled with the appropriate number of binary
collisions, and v2, the second moment of the Fourier decomposition
of the measured identified momentum spectra.

\begin{figure}[hbtl]
\hspace{0.2cm}
\includegraphics[width=2.9in, bb=0 0 400 400]{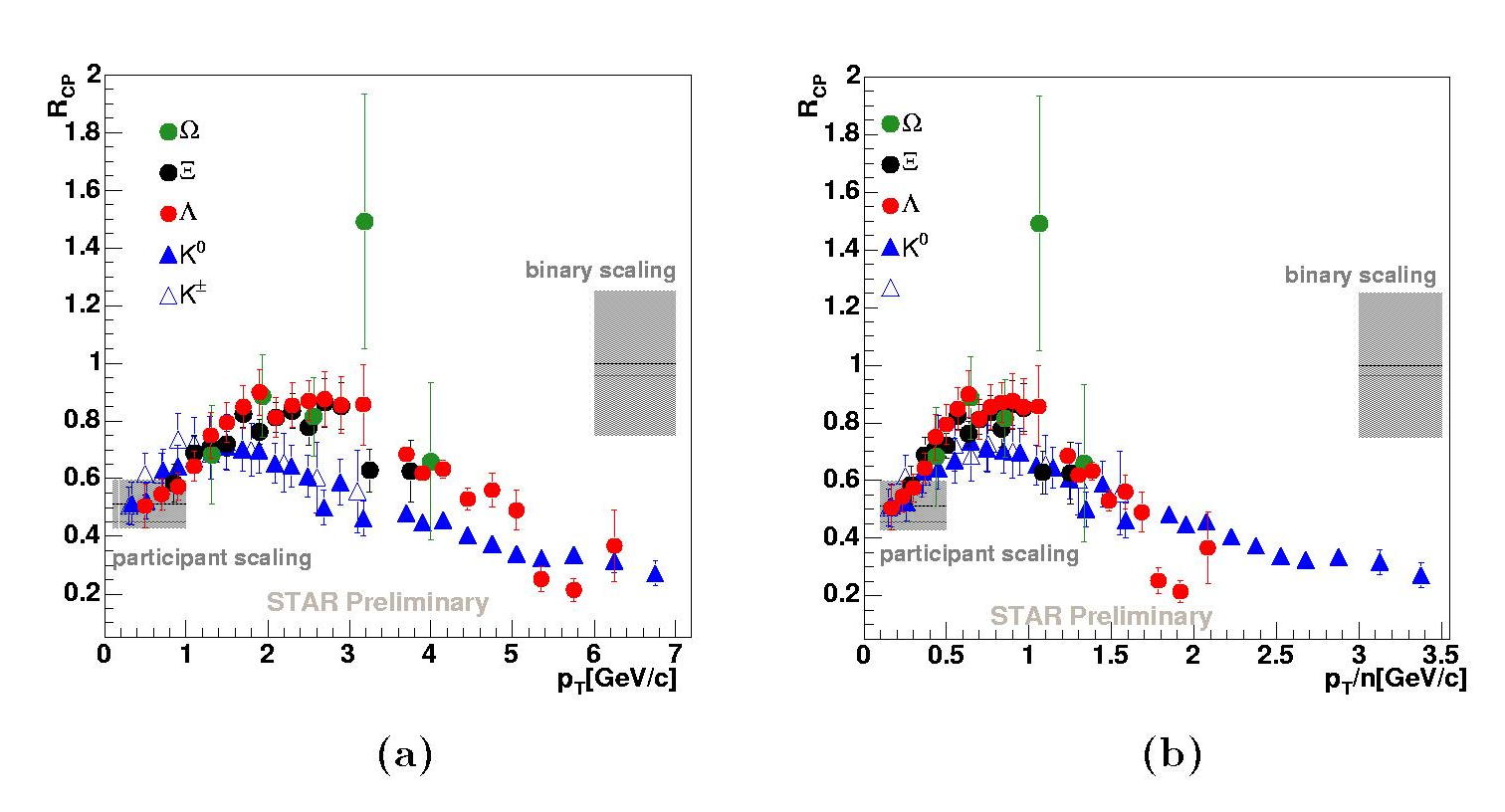} \label{fig:3}
\caption{STAR data on (a) R$_{CP}$ vs p$_{T}$ in Au+Au Collisions at
$\sqrt{s_{NN}}$ = 200 GeV. (b) R$_{CP}$ vs p$_{T}$/n  in Au+Au
collisions at $\sqrt{s_{NN}}$ = 200 GeV using n=3 for baryons and
n=2 for mesons. R$_{CP}$ is calculated from 0-5\% and 40-60\%
central Au+Au collisions.}
\end{figure}

\begin{figure}
\hspace{2.cm}
\includegraphics[width=2.9in, bb=0 0 400 400]{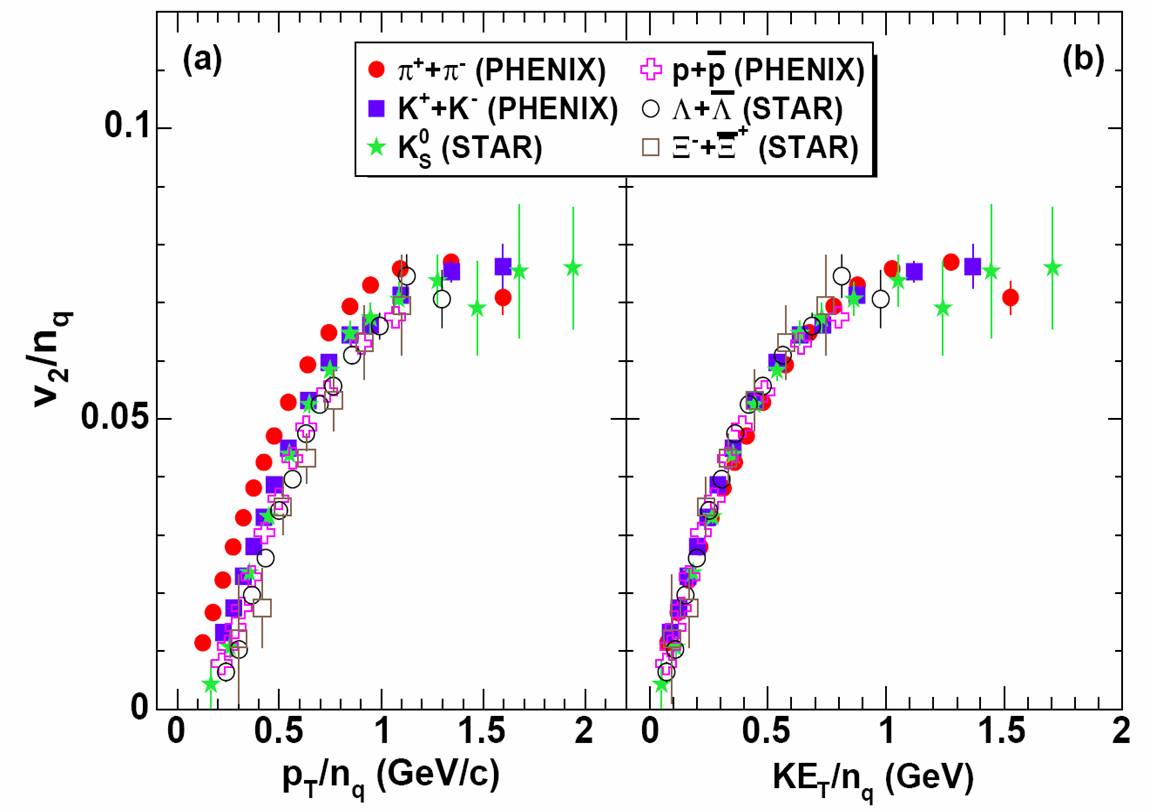} \label{fig:4}
\caption{ v2/n$_{q}$ vs. p$_{T}$/n$_{q}$ and KE$_{T}$/n$_{q}$ for
several particle species measured by STAR and PHENIX as indicated in
Au+Au collisions at $\sqrt{s_{NN}}$ = 200 GeV \cite{v2-summary}.}
\end{figure}

The scaling parameter could be called the number of constituent
quarks or the number of valence quarks. Quark scaling apparently
works, but the relevant degree of freedom is not well defined.
Popular recombination models \cite{bass, greco} have taken the
approach of using thermalized constituent quarks, with a well
defined mass, to describe not only the flow data but also the
unexpected baryon to meson (B/M) ratio at intermediate p$_{T}$. A
comparison to STAR B/M measurements is shown in Fig.4.

\begin{figure}
\hspace{1.cm}
\includegraphics[width=2.9in, bb=0 0 400 400]{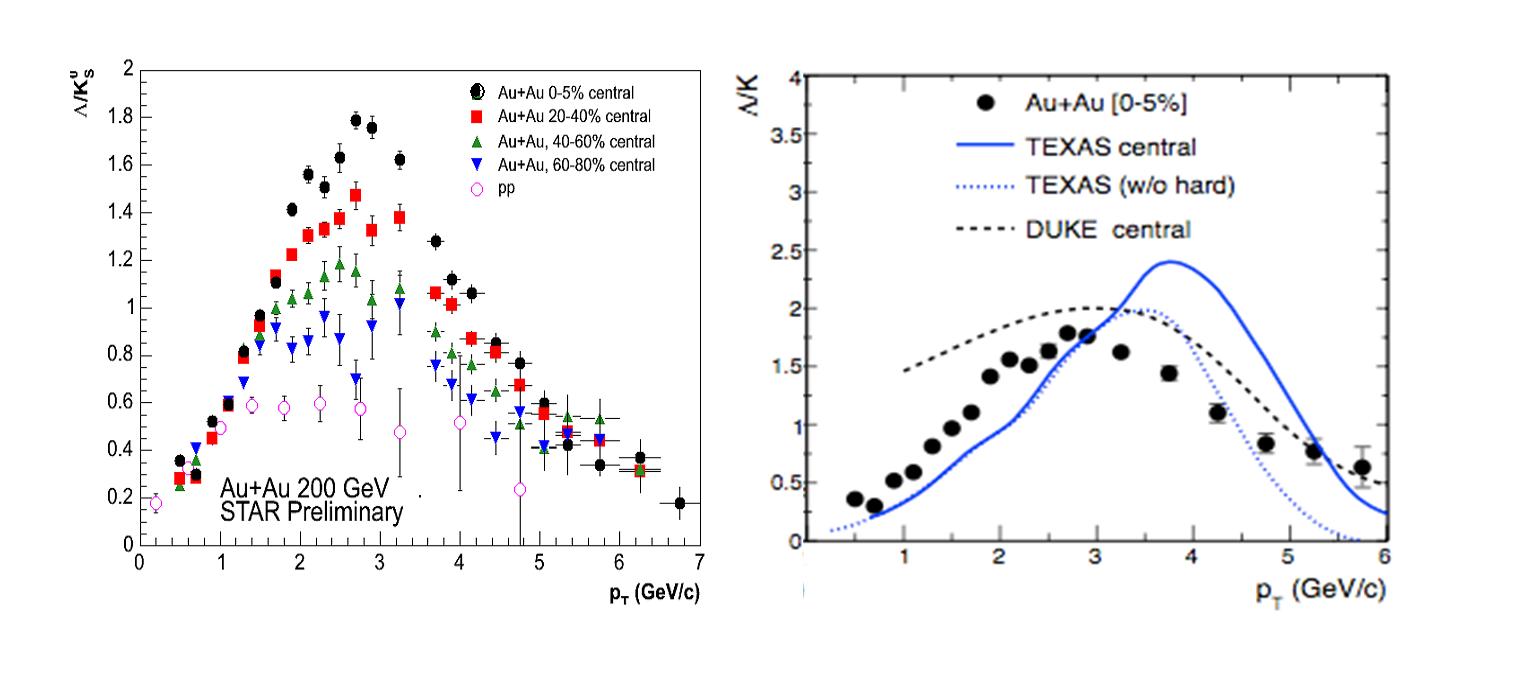} \label{fig:4}
\caption{a.) $\Lambda$/K$^{0}_{s}$ measurements in pp and AA
collisions at different centralities as measured by STAR. b.)
Comparison of central $\Lambda$/K$^{0}_{s}$ to recombination models
\cite{bass,greco}.}
\end{figure}

The main features of the central data results are rather well
described by the models. The B/M ratio peaks at about 2-3 GeV/c and
it is considerably higher than in proton proton collisions. It is
interesting to note, though, that the pp data already show a 'hump'
at intermediate p$_{T}$. This hump is rather well described by a
tuned PYTHIA fit \cite{mheinz}, which shows that string
fragmentation can describe the overall trend of the baryon to meson
production, but in terms of the amplitude of the B/M ratio any
fragmentation model underestimates the enhanced baryon production by
at least a factor 2-3. Recombination on the other hand seems to be
easily tuneable to the proper B/M ratio. The main reason is that in
a thermalized fireball it is easy to achieve large momentum overlap
of the quarks in the thermal pool, which then leads to the
production of higher momentum baryons, whereas the fragmentation of
very high momentum quarks into intermediate p$_{T}$ baryons is a
rare process. The required thermal pool of quarks is not achievable
in elementary collisions, but likely in heavy ion collisions.

A direct comparison of the spectral features obtained in pp and AA
collisions seems to further confirm the change in hadronization
mechanism, at least for hadrons in a specific momentum range. The
proton-proton measurements at RHIC reveal a breakdown of the
so-called m$_{T}$-scaling at intermediate p$_{T}$ \cite{star-str}.
Instead of a common scaling for all identified m$_{T}$ spectra,
which was established in ISR measurements at lower collision
energies, the RHIC data show a baryon/meson scaling at sufficiently
high transverse momentum. This can be explained by requiring
di-quark formation for baryon production, which leads to a di-quark
suppression factor which needs to be applied to the baryon spectra
in order to find a common hadron scaling. This effect is well
described by the gluon fragmentation model in PYTHIA \cite{mheinz}.
It is the first experimental evidence for di-quark formation at
RHIC, though. Di-quark formation leads to baryon/meson differences
but it can not describe the constituent quark scaling measured in AA
collisions. In fact di-quark formation should lead to a distinct
lack of scaling inasmuch as a diquark-quark based formation process
should scale similar to the quark-antiquark based process.
Investigations of scaling of balance functions for identified
particles in pp are underway to test this hypothesis. The validity
of the di-quark picture to describe the m$_{T}$-scaling of the
identified spectra in pp collisions requires gluon dominance in the
fragmentation process at RHIC energies. Besides the m$_{T}$ scaling,
the lack of discernible differences in the particle vs.
anti-particle production over the kinematic range measured at RHIC,
and the enhanced gluon fragmentation contribution in PYTHIA and
fragmentation function fits \cite{akk}, necessary to describe RHIC
data \cite{star-str,star-spec}, shows that at these collision
energies the parton interactions are indeed dominated by low x
gluons. This dominance is likely to further increase at the LHC due
to the even lower x coverage at the higher energies.

Extensions to the simple inclusive B/M ratio measurements in heavy
ion collisions have been recently performed by the STAR
collaboration. B/M ratios were measured in structures which appeared
in high momentum two-particle correlation measurements. Fig.5a shows
a comparison of B/M ratios in same-side and away-side jet cones
triggered by a high momentum charged particle \cite{zuo}, Fig.5b
shows a comparison of B/M ratios in the same side jet cone and the
same-side long-range correlation ridge as measured by STAR
\cite{bielcikova}. In both cases it seems that inside the unquenched
jet the B/M ratio is consistent with expectations from fragmentation
models, whereas the in-medium response to the traversing jets leads
to a ratio that is better described by the recombination scenario.

\begin{figure}
\vspace{-1.cm} \hspace{1.cm}
\includegraphics[width=3.in, bb=0 0 400 400]{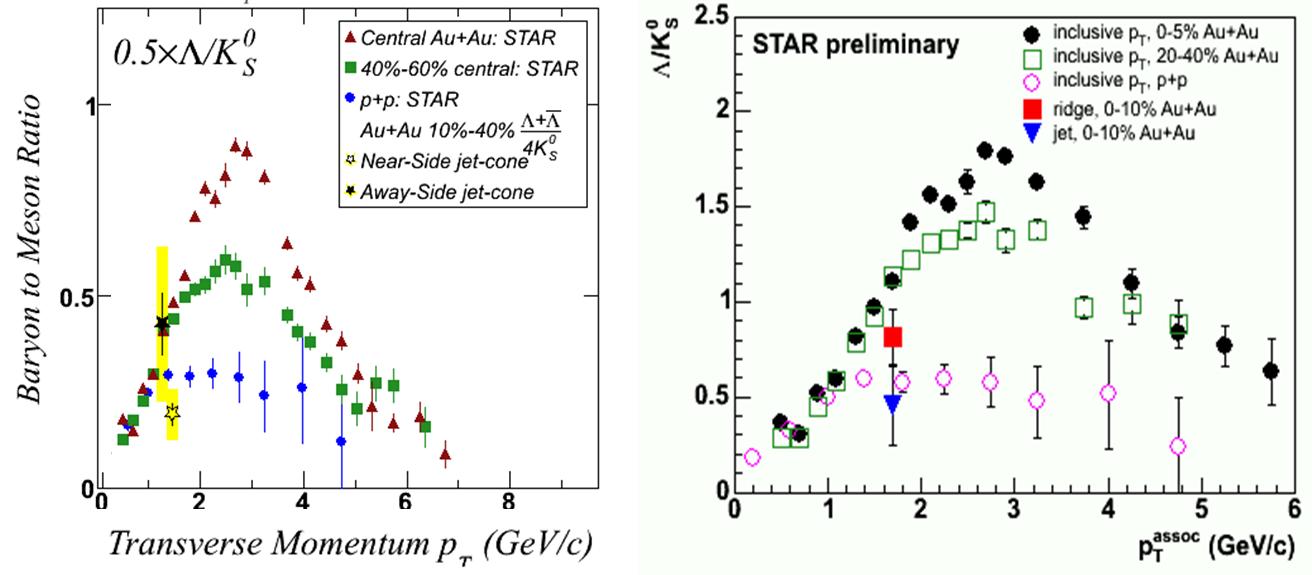} \label{fig:4}
\caption{ STAR measurements of the inclusive $\Lambda$/$K^{0}_{s}$
ratio as a function of centrality and transverse momentum, a.)
compared to ratios measured in the same-side and away-side jet
region from triggered two-particle correlations \cite{zuo}, b.)
compared to ratios measured in the same-side jet and the long-range
correlation ridge region from triggered two-particle correlations
\cite{bielcikova}.}
\end{figure}

At higher momenta (p$_{T}$ $>$ 6 GeV/c) the particle identified
spectra exhibit all the features of pure jet fragmentation, both in
single particle and particle correlation measurements
\cite{star-jets, phenix-jets}.

\section{Does strong coupling require a special degree of freedom ?}

It is interesting to note that there is a total lack of constituent
quark mass dependence in the scaling of v2 as shown in Fig.3. In
recombination approaches this is largely attributed to the fact that
the input constituent quark mass of the up, down and strange quarks
is quite similar (300 and 460 MeV, respectively) and that all
identified particles measured until recently did not include heavier
flavors. The recent measurement of the nuclear suppression factor
and the elliptic flow for D-mesons, based on electrons from the
semi-leptonic decay of the heavy mesons
\cite{heavy-data1,heavy-data2,heavy-data3}, allows us to determine
the applicability of partonic recombination a little further, and
early results seem to indicate that both, the R$_{AA}$ and the v2
measurements, can only be explained if one assumes identical
p$_{T}$-dependencies for the flow and the quenching of light and
heavy quarks as is shown in Fig.6 for R$_{AA}$ \cite{mischke} and in
Fig.7 for v2 \cite{v2-summary}.

\begin{figure}
\vspace{-1.5cm} \hspace{3.5cm}
\includegraphics[width=3.5in, bb=0 0 400 400]{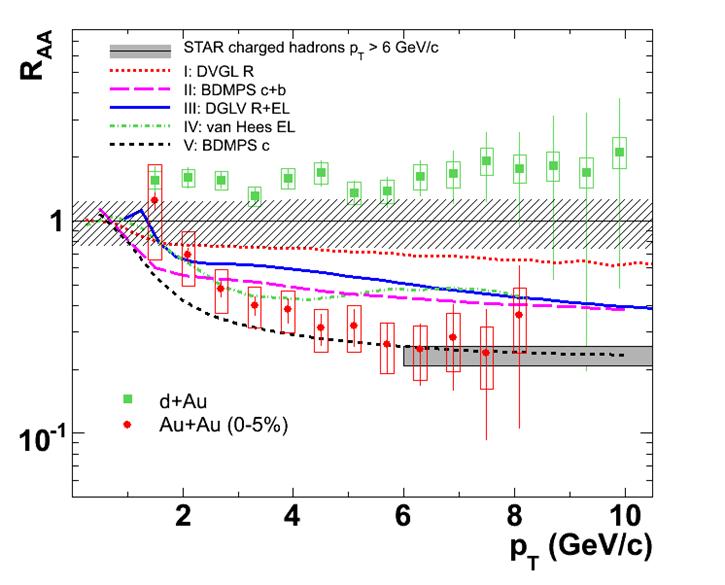} \label{fig:6}
\caption{Nuclear suppression factor for non-photonic single electron
spectra in semi-central Au+Au collisions at RHIC compared to the
R$_{AA}$ for charged hadrons (i.e. light quark suppression) and
various models \cite{mischke}.}
\end{figure}

\begin{figure}
\vspace{-1.5cm} \hspace{3.cm}
\includegraphics[width=3.in, bb= 0 0 400 400]{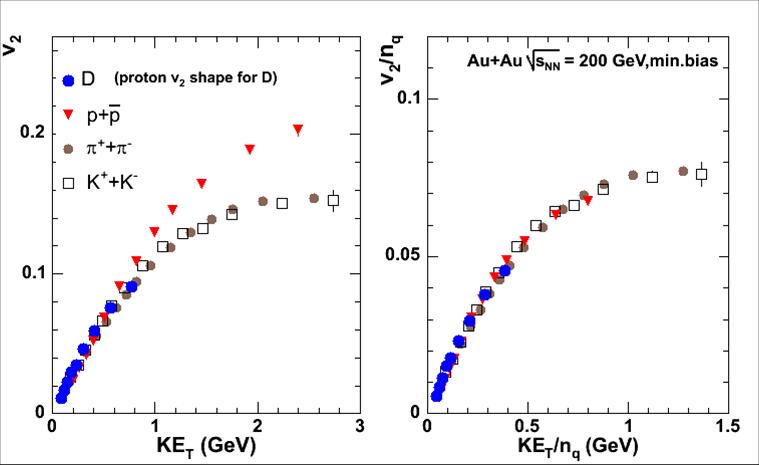} \label{fig:7}
\caption{Elliptic flow (non-scaled and n$_{q}$ scaled) for
non-photonic single electron spectra in semi-central Au+Au
collisions at RHIC compared to light quark hadrons
\cite{v2-summary}.}
\end{figure}

At some high momentum the mass of the bare or constituent heavy
quark should be negligible, but this should not be the case for the
intermediate momenta measured here. Many models, as shown in Fig.4,
have been proposed to address these measurements and in particular
the apparent lack of a dead cone effect for induced gluon radiation,
as well as the lack of a heavy quark mass dependence in the v2. The
most successful of these models try to give the heavy quark a
special status, by postulating either the survival of heavy quark
resonant states above T$_{c}$ \cite{rapp,rapp2} or the reduced
formation time of heavy quark hadrons from the partonic phase
\cite{vitev}. Fig.8 shows a comparison of the data to the heavy
quark bound state model.

\begin{figure}
\hspace{1.cm}
\includegraphics[width=2.0in, bb=0 0 400 400]{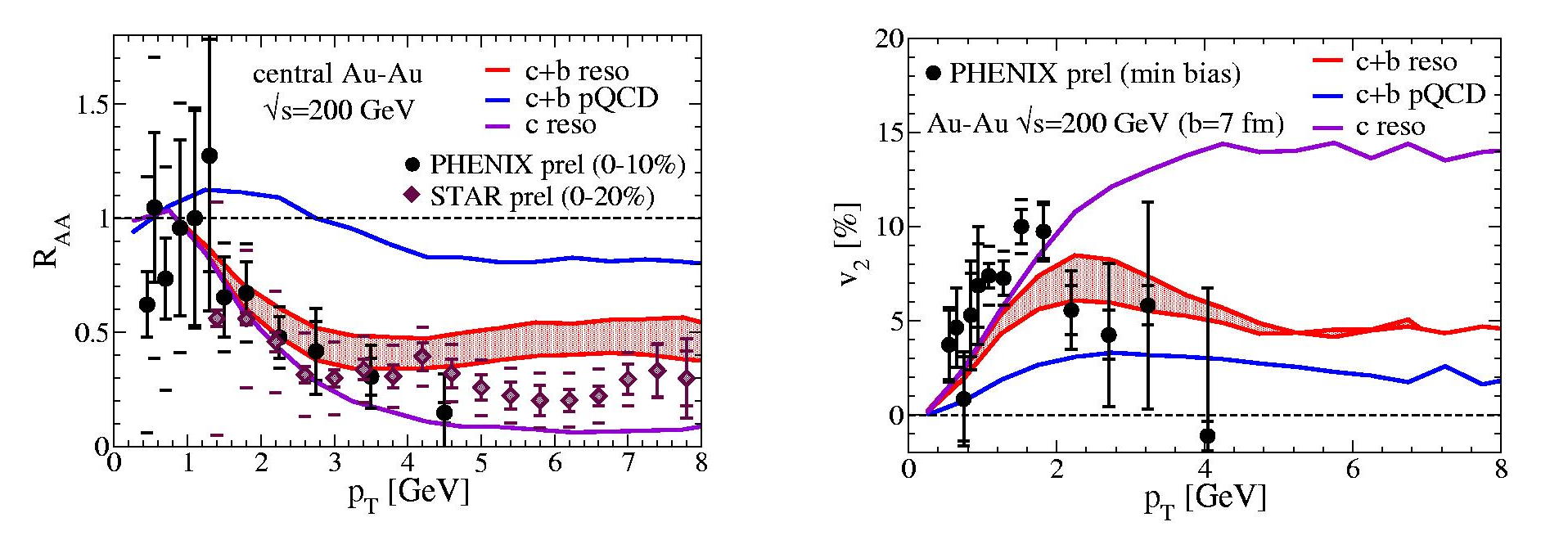} \label{fig:8}
\caption{Nuclear suppression factor (left panel) and elliptic flow
(right panel) for non-photonic single electron spectra in
semi-central Au+Au collisions at RHIC. Data
\cite{heavy-data1,heavy-data2,heavy-data3} are compared to theory
predictions \cite{rapp2} using Langevin simulations with elastic c-
and b-quark interactions in an expanding QGP fireball and
heavy-light quark coalescence at hadronization \cite{rapp}.}
\end{figure}

The near identical p$_{T}$-dependence of the v2 and the quark energy
loss between light and heavy quarks is very striking, though, and
might require a much more fundamental explanation. One possibility
is that the quasi-particle state formed near T$_{c}$ is really not
depending on the constituent or even bare quark mass concept, but
rather simply the number of partons, which could be mostly gluons,
until close to hadronization. Still, for a dynamic evolution measure
such as the v2 as a function of p$_{T}$, the dynamics of the degree
of freedom has to play a role, and at least the effect of the bare
quark mass should be measurable if we indeed probe the fragmentation
or recombination of quarks. A detailed measurement of reconstructed
D-mesons and B-mesons is sorely needed to remove the ambiguities in
the semi-leptonic measurements, and future measurements of high
momentum heavy flavor mesons and baryons should answer the question
whether the liquid phase above the critical temperature requires
indeed a special degree of freedom to describe all features of
hadronization from a dense deconfined medium.

\section{The relevance of RHIC results to the FAIR program}

Besides the very detailed and strong evidence for deconfinement at
RHIC energies, the data reveal a surprising lack of evidence for
chiral symmetry restoration. Vector meson and resonance measurements
have been performed to new levels of precision at RHIC, in
particular in the sector of heavy hadronic resonances, but the
measurements are mostly used to determine the lifetime of the
produced partonic and hadronic systems through detailed mapping of
re-interaction probabilities \cite{markert}. In these measurements
the properties of the resonances, such as mass, width, and branching
ratios are generally in very good agreement with the particle data
group references. There are small variations in mass and width for
certain resonances as a function of their momentum, but they have
been measured consistently in pp, dA, and AA collisions
\cite{fachini1} and therefore should not be attributed to chiral
symmetry restoration. Recently PHENIX has shown results that might
indicate behavior similar to the NA45 and NA60 low mass di-lepton
measurements \cite{phenix-dil}, but whether this is evidence for
medium modification of vector mesons remains to be seen. It seems
that either our measurements are not sensitive to chiral symmetry
restoration or that the chiral transition might indeed decouple from
deconfinement, which is in disagreement with lattice QCD
calculations. A very high luminosity program at FAIR should enable
more detailed measurements of medium modification, in particular for
chiral partners in the heavy quark sector. It is remarkable to
realize that although open charm production is at threshold at FAIR
energies, the yield of open charm obtained in a 25 week run at CBM
is about an order magnitude larger than the yield STAR obtains over
the same period of time \cite{senger}. Detailed measurements, not
only of chiral partners, but also particle identified elliptic flow,
radial flow and jet quenching might therefore be possible albeit at
a slightly lower p$_{T}$-range. The main purpose of these
measurements should be to map out the disappearance of the strong
sQGP signatures, such as quark scaling, hydro scaling, and high
p$_{T}$ suppression. These results will complement the thrust of the
CBM program, which emphasizes the search for a critical point in the
QCD phase diagram.

\vfill\eject

\end{document}